\documentclass[fleqn,usenatbib]{mnras}

\usepackage[T1]{fontenc}

\DeclareRobustCommand{\VAN}[3]{#2}
\let\VANthebibliography\thebibliography
\def\thebibliography{\DeclareRobustCommand{\VAN}[3]{##3}\VANthebibliography}

\usepackage{graphicx}	% Including figure files
\usepackage{amsmath}	% Advanced maths commands
\usepackage{newtxtext,newtxmath}
\title{Pulsar Observations at Low Frequencies: Applications to Pulsar Timing and Solar Wind Models}

\author[P. Kumar et al.]{
P. Kumar,$^{1}$ %\thanks{E-mail: mn@ras.org.uk (KTS)}
S. M. White,$^{1,2}$
K. Stovall,$^{1}$
J. Dowell$^{1}$
and G. B. Taylor$^{1}$
\\
$^{1}$Department of Physics and Astronomy, University of New Mexico, 210 Yale Blvd NE, Albuquerque, NM 87106, USA\\
$^{2}$Space Vehicles Directorate, Air Force Research Lab,  Kirtland AFB, NM\\
}

\date{Accepted XXX. Received YYY; in original form ZZZ}

\pubyear{2021}

\begin{document}
\label{firstpage}
\pagerange{\pageref{firstpage}--\pageref{lastpage}}
\maketitle

\begin{abstract}
 Efforts are underway to use high-precision timing of pulsars in order to detect low-frequency gravitational waves. A limit to this technique is the timing noise generated by dispersion in the plasma along the line of sight to the pulsar, including the solar wind. The effects due to the solar wind vary with time, influenced by the change in solar activity on different time scales, ranging up to $\sim 11$ years for a solar cycle. The solar wind contribution depends strongly on the angle between the pulsar line of sight and the solar disk, and is a dominant effect at small separations. Although solar wind models to mitigate these effects do exist, they do not account for all the effects of the solar wind and its temporal changes. Since low-frequency pulsar observations are most sensitive to these dispersive delays, they are most suited to test the efficacy of these models and identify alternative approaches. Here, we investigate the efficacy of some solar wind models commonly used in pulsar timing using long-term, high-cadence data on 6 pulsars taken with the Long Wavelength Array, and compare them with an operational solar wind model. Our results show that stationary models of the solar wind correction are insufficient to achieve the timing noise desired by pulsar timing experiments, and we need to use non-stationary models, which are informed by other solar wind observations, to obtain accurate timing residuals.
\end{abstract}


%%%%%%%%%%%%%%%%%%%%%%%%%%%%%%%%%%%%%%%%%%%%%%%%%%

%%%%%%%%%%%%%%%%% BODY OF PAPER %%%%%%%%%%%%%%%%%%

\section{Introduction} \label{sec:intro}
Pulsars are rapidly rotating and highly magnetized neutron stars (\citealt{1968Natur.217..709H}), which produce beamed broadband emission, due to relativistic effects near the neutron star, and are most commonly observed at radio frequencies. This emission appears as a pulsed signal due to the very stable rotation periods of neutron stars, producing a pulse that is only a fraction of a period in width when the cone of emission from the pulsar sweeps across the line-of-sight (LoS) to the observer. Even though individual pulses from a pulsar vary, the average pulse profile is generally stable in frequency over timescales of years to decades \cite[]{1975ApJ...198..661H,10.1111/j.1365-2966.2011.20041.x}. This unique property of pulsars allows them to serve as precision clocks, where the arrival time of a subsequent pulse can be predicted to high accuracy.

Among the population of pulsars, millisecond pulsars \cite[MSPs; see, e.g.,][]{Backer1982} are a sub-class that are believed to be older pulsars, which are ``recycled" by accreting material from a companion and thereby have been spun up to millisecond rotational periods. MSPs can provide precision in the period measurement of one part in $10^{13}$ or better, compared to one part in $10^7$ for a normal pulsar \cite[see, e.g.,][]{Lorimer2008}. This makes them useful for a range of projects such as tests of general relativity, navigation, etc., via pulsar timing experiments \cite[see, e.g.,][]{2017JPhCS.932a2002M}. Perhaps the most interesting among these is the long-term monitoring of a set of high-precision pulsars to detect gravitational waves from supermassive blackhole binaries in the nano-hertz regime \citep[e.g.,][]{2013ApJ...762...94D}. These experiments are generally referred to as pulsar timing arrays (PTAs), with the North American Nanohertz Observatory for Gravitational Waves \cite[NanoGrav;][]{nanograv}, European Pulsar Timing Array \cite[EPTA;][]{epta}, and Parkes Pulsar Timing Array \cite[PPTA;][]{ppta} being three major efforts in this direction.

Even though pulsars intrinsically provide a highly stable pulse, the intervening medium between the pulsar and the Earth induces effects in the propagation of pulsar radiation, which vary on a range of timescales \cite[see, e.g.,][]{10.1093/mnras/sts486}. As the radiation passes through the intervening turbulent ionized medium, scintillation, scattering, dispersive effects, and Faraday rotation are imposed on the signal. Among these, perhaps the easiest and most important to correct is the dispersive delay due to the ionized interstellar medium (IISM), the solar wind (SW), and Earth's ionosphere, each introducing a dispersive time delay $\Delta t$, along the LoS. The effective time delay due to these effects is proportional to the LoS electron column density and the inverse-square of the observed frequency $f$ (in MHz), as given by the relation,
\vspace{-0.3cm}
\begin{equation} \label{eq:1}
\hspace{3cm}    \Delta t = K \frac{DM}{f^{2}}
\end{equation}

\noindent where $K$=4.15 $\times$ 10$^6$ ms \cite[see,][]{2004hpa..book.....L} and DM, usually expressed in pc cm$^{-3}$, is the integrated LoS free electron density, referred as the \textit{dispersion measure}, given by the relation,

\begin{equation} \label{eq:2}
\hspace{2.8cm}    DM =  \int_{0}^{d} n_e \,dl 
\end{equation}  

\noindent where $d$ is the distance to the pulsar from Earth, and $n_e$ is the electron number density. Since the contribution from each of the three terms (IISM, SW, and terrestrial ionosphere) creates a variation in the measured DM with time, due to changes in the LoS electron density structure in each medium, they need to be accounted for via complementary observation and modeling to reduce the error in pulsar timing measurements that may arise from insufficient dispersion correction. To achieve their goal, PTAs need to achieve a timing accuracy of about 100 ns, corresponding to a DM uncertainty of the order of $~5 \times 10^{-5}$
pc cm$^{-3}$. Since sources other than those discussed here may also contribute to the timing budget, it is desirable that the target for correction of SW and IISM effects be lower than this. The most recent estimates of timing noise for PTAs are of the order of 1--10 $\mu s$ \cite[see, e.g.,][]{2021ApJS..252....5A,2020PASA...37...20K,2016MNRAS.458.3341D}. As indicated by Equation \ref{eq:1}, studies of efforts to mitigate dispersive delays are best done at lower frequencies where the impact of dispersion is greater. Low-frequency pulsar observations are providing insights into the distribution of electrons in the IISM \cite[see, e.g.,][]{2017ApJ...846..104K,2019ApJ...875..146B} and constraints on ionospheric electron density measurements \cite[see, e.g.,][]{Malins2018}, among others. Recent low-frequency studies on the effect of the IISM and SW on pulsar DM variations \cite[see, e.g.,][]{Tiburzi2019,Tiburzi2021} suggest that the existing SW models for pulsar timing experiments are inadequate to make significant improvements to timing noise levels.

Since the DM contribution from the ionosphere is of the order $10^{-5}$ pc cm$^{-3}$, it is much smaller than the current sensitivities of PTAs, and hence we do not address the ionospheric contribution in this article. The next significant contribution to DM variations comes from the SW. This is of the order of $10^{-3} - 10^{-4}$ pc cm$^{-3}$, depending on the solar elongation of the pulsar, and so needs to be accounted for \cite[see, e.g.,][]{Tiburzi2021}. Most pulsar timing analysis packages use a simple spherical model of the SW. The SW itself is composed of slow dense streams and faster, more diffuse streams, and it is necessary to understand the mix of such structures along the LoS to the pulsar in order to make the appropriate corrections. Additionally, there are transient phenomena called coronal mass ejections (CMEs) associated with eruptions on the Sun, which are harder to model but contribute significantly to the SW DM \cite[see, e.g.,][]{shaifullah, kkumar}. Temporal variations on many different timescales also occur in the SW due to solar activity: timescales of days (active region evolution), the solar rotation period (27 days), and the solar cycle ($\sim$11 years). As mentioned before, solar elongation plays a critical role in the size of the correction needed due to the much higher densities occurring close to the Sun.

The largest contribution to the variation can arise from fluctuations in the IISM, which can be of the order of $10^{-3}$ pc cm$^{-3}$. Previous studies have shown significant change over periods of a few days to weeks \citep[e.g.,][]{1993ApJ...404..636B}; \citet{2004MNRAS.353.1311H} proposed the relation $|\frac{d(DM)}{dt}|  \approx 0.0002 \sqrt{DM}$ pc cm$^{-3}$ yr$^{-1}$, relative to typical DM values of tens of pc cm$^{-3}$. The DM fluctuations arise mainly due to relative motion of spatial structures in the IISM that occur along the LoS to the pulsar. The IISM can have small-scale structures which drive fluctuations \cite[see, e.g.,][]{2020ApJ...892...26B}. Finally, at times it can be difficult to disentangle the effects due to the SW and the IISM \cite[see, e.g.,][]{Tiburzi2019}, which makes it difficult to create a model applicable in all situations.

In this article, we use long-term regular monitoring observations of pulsars with the Long Wavelength Array (LWA) at frequencies below 88 MHz, along with data from several higher-cadence solar campaigns, to study DM fluctuations due to the SW and IISM and assess the efficacy of existing SW models for PTA experiments, including comparison with a non-stationary model of the solar wind. Section 2 describes the observational and data acquisition setup, and section 3 describes the SW and relevant models evaluated. Section 4 describes the methods used to calculate the DM via pulsar timing, calculation of the SW contribution from models, and the modeling of the IISM contribution. Section 5 and 6 present relevant results and discussion, respectively. Finally, we conclude in section 7.

\section{Observations} \label{sec:observe}
The LWA is a low-frequency radio telescope array \cite[see,][]{2012JAI.....150004T,2013ITAP...61.2540E}. Currently, there are two stations, LWA1, located near the Karl G. Jansky Very Large Array west of Socorro in central New Mexico, and LWA-SV, located on the Sevilleta National Wildlife Refuge north of Socorro. Each station consists of 256 dual-polarization dipole antennas that operate between 3 and 88 MHz. Pulsar observations are one of the main science areas of the first station, LWA1, and regular monitoring of pulsars is being done to study pulsar properties over time as well as those effects that are dominant at low radio frequencies such as dispersion and scattering \cite[see, e.g.,][]{2019ApJ...875..146B}. The regular monitoring of pulsars is done at a $\sim$ 3-week cadence, depending on resource availability, and the reduced data products are stored on a public archive\footnote{LWA Pulsar Archive: \url{https://lda10g.alliance.unm.edu/PulsarArchive/}}. For details of observation and basic data reduction for the monitoring program, see \citet{2015ApJ...808..156S}. The current monitoring program observes over 100 pulsars, for some of which data is available since 2012. Apart from the regular monitoring, we also perform campaigns during the solar transit of some pulsars at a higher cadence, to capture the effects of the SW on pulsar properties.

LWA1 has four independent delay-and-sum beams, formed by summing the individual dipoles of the array after applying appropriate delays to each element, of which two are used for the pulsar monitoring observations at a given time. Each beam can have two spectral tunings with independent center frequencies, each with a maximum bandwidth of 19.6 MHz and dual polarization. Pulsar monitoring observations are done using central frequencies of 35.1 MHz and 49.8 MHz for one beam, each with 19.6 MHz bandwidth and dual-polarization, and similarly with 64.5 MHz and 79.2 MHz central frequencies for the other beam, simultaneously. The raw data is then reduced using an automated data reduction pipeline\footnote{\url{https://github.com/lwa-project/pulsar}} \footnote{\url{https://github.com/lwa-project/pulsar_archive_pipeline}}, which uses tools from a combination of standard pulsar data reduction software such as TEMPO, PSRCHIVE\footnote{\url{http://psrchive.sourceforge.net/}} (\citealt{2012AR&T....9..237V}), DSPSR\footnote{\url{http://dspsr.sourceforge.net/}}, PRESTO\footnote{\url{https://github.com/scottransom/presto}} and the LWA Software Library\footnote{\url{https://github.com/lwa-project/lsl}} (\citealt{2012JAI.....150006D}). This gives us 4 sets of independent data products, which includes psrfits archives, one for each tuning, plus one after carefully combining them, taking into account the filter roll-off at the edges, which are then stored in the LWA Pulsar archive.
\begin{table*}
    \begin{center}
    \tabcolsep=0.01cm
    \begin{tabular}{cccccccc}
    \\[-1.8ex]\hline 
    \hline \\[-1.8ex]
    {PSR Name} & \hspace{0.5cm}{\# Phase} & \hspace{0.5cm}{\# Channels} & \hspace{0.5cm}Period & \hspace{0.5cm}DM & \hspace{0.5cm}Ecliptic & {Time-span} & {\# Epochs} \\
      & \hspace{0.5cm}bins &  & \hspace{0.5cm}(ms) & \hspace{0.5cm}{(pc cm$^{-3}$)} & \hspace{0.5cm}{Latitude (deg)} &  &  \\
    \hline
    J0030+0451&\hspace{0.5cm}256&256&\hspace{0.2cm}4.87&\hspace{0.5cm}4.332681$\pm$0.000057&1.45&\hspace{0.5cm}2013-12 to 2021-04&92\\
    \hline
    J0034-0534&\hspace{0.5cm}128&128&\hspace{0.2cm}1.88&\hspace{0.5cm}13.765199$\pm$0.000007&-8.53&\hspace{0.5cm}2013-12 to 2021-04&70\\
    \hline
    B0950+08&\hspace{0.5cm}1024&4096&\hspace{0.2cm}253.07&\hspace{0.5cm}2.96927$\pm$0.000080&-4.62&\hspace{0.5cm}2015-08 to 2021-03&101\\
    \hline
    B1257+12&\hspace{0.5cm}256&256&\hspace{0.2cm}6.22&\hspace{0.5cm}10.153322$\pm$0.000014&17.58&\hspace{0.5cm}2016-09 to 2021-05&44\\
    \hline
    J1400-1431&\hspace{0.5cm}256&128&\hspace{0.2cm}3.08&\hspace{0.5cm}4.9333$\pm$0.000027&-2.11&\hspace{0.5cm}2016-03 to 2021-04&39\\
    \hline
    J2145-0750&\hspace{0.5cm}512&256&\hspace{0.2cm}16.05&\hspace{0.5cm}9.004347$\pm$0.000034&5.31&\hspace{0.5cm}2013-12 to 2021-03&79\\
    \hline
    \end{tabular}
    \end{center}
    \caption{Pipeline reduced number of phase bins and channels of archive files for the six analyzed pulsars. The other columns are the values of pulsar spin period, the used DM for calculation, the span of data used for each pulsar, and the number of epochs available in the end (after all bad observations are removed).}
    \label{tab:1}
\end{table*}

In our current study, we work with a set of 5 MSPs, J0030+0451, J1400-1431, J2145-0750, J0034-0534, B1257+12, and the slow pulsar B0950+08, all of which are close to the ecliptic plane (ecliptic latitudes between -9\degr\ and +18\degr). The pulsars were chosen from the LWA pulsar database based on their spin period and closeness to the ecliptic, all below 20 degrees, keeping only those MSPs for the final sample, for which we can obtain good timing solutions with the existing data. This was intended to show the effect of SW on millisecond class pulsars, relevant for pulsar timing experiments. Additionally, PSR B0950+08, being very bright in the LWA band, was selected as a test pulsar. We use all the data available to us for these pulsars from the LWA monitoring program and solar campaigns. Each of the MSPs is observed for a duration of 2 hr at each epoch, while B0950+08, bright ($>25$ sigma detection) at low frequencies, is observed for 0.5 hrs, to get sufficient signal-to-noise. Following the initial pipeline reduction, the stored archive file has the number of channels and number of phase bins as stated in Table \ref{tab:1}, each with 30 s sub-integration time.

\vspace{-0.25 cm}
\section{Solar Wind Models}\label{sw}
The solar wind is a stream of magnetized plasma and charged particle emission from the Sun driven outwards by the pressure of the hot solar corona. The SW has been studied for a long time using spacecraft observations of the Sun and ground-based observations, mainly in the radio frequency regime. The SW exhibits the Sun's activity cycle of 11 years, as well as variability on shorter timescales such as the 27-day rotation period, and needs to be accounted for either via modeling or regular observations. The density in the SW falls off with radial distance, and thus effects are strong near the Sun and diminish at larger angular separations. Observations via the Helios and Ulysses spacecraft provided evidence for the bimodal nature of the SW, which becomes clearer at large solar latitudes \cite[see, e.g.,][]{1996Ap&SS.243...87C}. A continuous change in velocity has been observed, where the two phases are (1) a slower and denser mode ($\sim$ 200 km s$^{-1}$), called the ``\textsc{slow wind}", and (2) a higher-velocity ($\sim$ 700 km s$^{-1}$) lower-density phase, called the ``\textsc{ fast wind}" \cite[see, e.g.,][]{Pierrard2020}. As shown by out-of-ecliptic measurements by the Ulysses spacecraft, the fast wind originates from coronal holes \cite[see, e.g.,][]{McComas1998}. In contrast, the source of slow wind is not clear due to different possible scenarios. As the streams propagate outwards, fast-moving streams can overtake a previous slow stream, creating compression regions called ``stream interaction regions'' \cite[SIRs;][]{Allen_2020}. These can be short-lived, but also may persist for long periods of one or more corotations of the Sun (``corotating interaction regions'', or CIRs). Both such regions lead to inhomogeneities, which will contribute to the DM measured from Earth. Other than these temporal and spatial variations, the Sun can also erupt, resulting in CMEs that can have high densities and speeds significantly faster than the fast streams, with occurrence rates varying between solar minimum and maximum \cite[see, e.g.,][]{Richardson2012}. The chance of a CME being encountered during a 1-2 hr pulsar observation is minimal but still not negligible, especially near the peak of solar activity. Such phenomena are difficult to model and will further contribute to the DM on lines of sight through the CME. 

\subsection{Common SW models for PTAs}
The simplest SW model assumes a constant speed of the wind preserving mass, which describes a totally spherical SW. The electron density contribution of this model to the DM decreases as the square of the radial distance from the Sun and the density profile is given by the relation,

\begin{equation}\label{eq:3}
\hspace{3cm} n_e(R) = \frac{n_0}{R^2}
\end{equation}

\noindent where $R$ is the radial distance from the Sun to the astronomical object, in units of AU and $n_0$ is the free electron density at Earth in units of cm$^{-3}$. The most commonly used pulsar timing packages TEMPO\footnote{\url{https://sourceforge.net/projects/tempo/}} (hereafter T1) and TEMPO2\footnote{\url{http://tempo2.sourceforge.net/}} \cite[(hereafter T2)]{Hobbs2006MNRAS.369..655H,Ed2006MNRAS.372.1549E} use this as the default model. The only difference between the two is that the former uses a default value of $n_0=10$, corresponding to a slow SW, whereas the latter uses $n_0=4$, assuming it to be fast. In general, the value assumed by T1 has been found to predict DM contributions larger than measurements suggest, whereas those of T2 are more reasonable, as shown by the prevalence of fast winds in solar wind observations, except during solar maximum \cite[see, e.g.,][]{Tokumaru}. Although these are the defaults, $n_0$ is a tunable parameter that can be adjusted by the user to better fit a particular data set, as shown by the authors in \citet{madison}, where they find a best-fit estimate of 7.9$\pm$0.2 cm$^{-3}$ for their data. As expected, any correction provided by using this value of $n_0$ will be bounded by the T1 and T2 models, and will likely be closer to that of former. Hence, we don't discuss this model explicitly. As is evident from Equation \ref{eq:3}, these spherical SW models only allow for a radial gradient correction while no other spatial structure of the SW is captured. Thus a spherical-wind model cannot represent density fluctuations due to the interaction of slow and fast phases of the SW, nor can it correct for any transient phenomena.

The other commonly-used SW models are those which describe the slow and fast phases of the SW independently. This can capture the large scale structure of the SW and dependence on the radial distance from the Sun. The electron number density in the fast wind phase is described by the following relation (hereafter \textsc{fast}), which is modeled after results from \citet[]{Guhathakurta1995,Guhathakurta_1998}, 

\begin{equation} \label{eq:4}
\hspace{1.5cm}\begin{split}
    n_e(R) = & 1.155\times 10^{11} R^{-2} + 32.3 \times 10^{11} R^{-4.39} \\ & + 3254 \times 10^{11} R^{-16.25} \mbox{~m}^{-3}
\end{split}
\end{equation}

\noindent where $R$ is the distance in solar radii units. Similarly, the slow wind is modeled using a combination of parameterization from \citet{muhle1981ApJ...247.1093M} and \citet{allen1947MNRAS.107..426A}, given by the relation (hereafter \textsc{slow}),

\begin{equation} \label{eq:5}
\hspace{1.5cm} \begin{split}
    n_e(R) = & 2.99\times 10^{14} R^{-16} + 1.5 \times 10^{14} R^{-6} \\ & + 4.1 \times 10^{11} (R^{-2} + 5.74 R^{-2.7}) \mbox{~m}^{-3}
\end{split}
\end{equation}

A combination of the \textsc{slow} and \textsc{fast} models has also been presented in \citet{y07} and implemented in the TEMPO2 software package. They fix the latitude range around the neutral field line and SW speed for both the modes to be $20$ degrees and $400$ km s$^{-1}$ respectively. The LoS is divided into segments that project to $5$ degrees at the solar surface, and using the propagation speed mentioned above as well as the heliographic coordinates of neutral field lines from the Wilcox Solar Observatory (WSO\footnote{\url{http://wso.stanford.edu/forms/prsyn.html}}) synoptic charts, they find the appropriate phase of the SW which affects that segment. After that, all the segments are summed over to get the total contribution. In general, this is a linear combination of the \textsc{slow} and \textsc{fast} models, and so any prediction by this method will be bounded by those two models, and we treat them separately in our analysis.

\subsection{The WSA-ENLIL model}
The Wang-Sheeley-Arge–ENLIL (WSA-ENLIL; hereafter \textsc{wsa}) SW model is a large-scale heliospheric model, serving as the NOAA operational model for forecasting SW conditions at Earth \citep{PMP11}\footnote{\url{https://www.swpc.noaa.gov/products/wsa-enlil-solar-wind-prediction}}. It has been well validated: of 15 models investigated by \citet{2015SpWea..13..316J}, WSA-ENLIL was second in its performance for predicting SW density at Earth. The model is driven by observations of the solar magnetic field B at the Sun's surface, and combines an empirical approach to determine conditions at the base of the SW \citep[see, e.g.,][and references therein]{2020SpWea..1802464M} with a 3D magnetohydrodynamic (MHD) code (ENLIL) that propagates the wind (and any CMEs present in the wind) out to Earth and beyond. Photospheric B measurements are extrapolated outwards to a potential-field source surface (where open magnetic field lines become radial) that is effectively the outer boundary of any closed field lines in the corona. SW speeds at the source surface (usually taken to be at 2.5 R$_{\sun}$) are empirically correlated with the degree of expansion of the corresponding open magnetic field lines. At this point the model is coupled to the \citet{Sch72} coronal current sheet model, and the MHD code ENLIL takes over at 21.5 R$_{\sun}$ to propagate the wind further outwards.

NOAA's Space Weather Prediction Center (SWPC) runs WSA-ENLIL on a daily basis to predict SW conditions. When significant CMEs occur, their physical properties are fitted to a cone model \citep{PMP11} and they are injected into the simulation: this can result in several additional runs per day as CME measurements are refined. The SW conditions generated by WSA-ENLIL depend critically on the input magnetic model. Photospheric measurements of B may be obtained from sources such as GONG\footnote{Global Oscillations Network, \url{https://gong.nso.edu/}}, SOLIS\footnote{Synoptic Optical Long-term Investigations of the Sun, \url{https://nso.edu/telescopes/nisp/solis/}}, and HMI\footnote{Helioseismic and Magnetic Imager on NASA's Solar Dynamics Observatory, \url{http://hmi.stanford.edu/}}. These measurements can be assimilated into a global magnetic field model, such as ADAPT\footnote{Air Force Data Assimilative Photospheric Flux Transport model} that uses a magnetic flux transport model to deal with the time evolution of unobserved fields such as those on the far side of the Sun \citep[e.g.,][]{2015SoPh..290.1105H}. While these approaches are being validated, the operational version of WSA-ENLIL continues to use a standard synoptic map generated from daily measurements near central meridian, with no correction for time evolution. 

The daily SWPC WSA-ENLIL runs are archived at \url{https://www.ngdc.noaa.gov/enlil/}, starting on 2013 November 19 (but with some gaps in coverage). These runs are designed to reproduce the time-varying behavior of the SW at Earth, and each contains 7 days of data at 1-hour time resolution, starting 2 days before the date on which it was generated. Since the full 3D global simulations with this time coverage would represent very large files, these archive files contain a limited representation of the full simulations. There are time series of density, velocity and radial magnetic field orientation specifically at Earth (and potentially other locations of interest, such as the spacecraft STEREO A and B). The simulations use radius, colatitude and longitude as variables. The archived files contain three 2D cuts through the full 3D simulation: a partial radius-latitude cut orthogonal to the ecliptic along the Sun-Earth line, for ecliptic latitudes $\pm60\degr$ from the Sun; the spherical surface at 1 AU; and, more useful for our purposes, the full ecliptic plane from 0.1 to 1.7 AU (the data are the output from the ENLIL calculation, which starts at 21.5 R$_{\sun}$ = 0.1 AU). Thus these archived files are only useful for lines of sight close to the ecliptic, but have the advantage that they are already available for nearly all of the hundreds of days for which we have pulsar data. For comparison with the pulsar DM measurements, we chose the archive file generated on the date closest to the day of observation (usually the same day), and picked the hourly ecliptic cut corresponding to the time of transit of the pulsar. The radius-longitude representation in the archive data (with resolution 0.003125 AU in radius and $2\degr$ in longitude) was regridded to a rectangular array of resolution 0.002 AU (0.43 R$_{\sun}$) for convenience.

The ecliptic longitude of the pulsar on the day of observation was determined, and the DM contribution from the WSA-ENLIL model was calculated by integrating from Earth through the model along the appropriate longitude. Since the archive files have no modelled data inside 0.1 AU, lines of sight that pass within $10\degr$ of the Sun cannot be calculated reliably. The density in the SW naturally peaks close to the Sun, so the main contribution to the DM for lines of sight within about $30\degr$ is well captured by the simulation extending out to 1.7 AU (i.e., the integration from Earth to the edge of the model covers up to 2.7 AU of distance). However, for lines of sight at larger angles from the Sun, the relative contribution to the DM from the SW beyond the radius of 1.7 AU can be significant. These contributions are calculated separately and added to the DM, for all solar elongations for which we evaluate the  \textsc{wsa} model. To handle this, we take the density at the outer edge of the model along the chosen line of sight, and assume that it drops as 1/distance$^2$ beyond that point.  With this assumption, the contribution from beyond 1.7 AU can be calculated analytically and shown to be

\begin{equation}
 \Delta DM\,=\,1 {\rm AU}\, *\,\frac{n_1 r_1^2} {\sin{\lambda_e}}\,\biggl(\frac{\pi}{2}\,-\,\arctan\frac{\sqrt{r_1^2\,-\,\sin^2{\lambda_e}}} { \sin{\lambda_e}}\biggr) 
\end{equation}
\newline
\noindent where $n_1$ is the electron density at the outer edge of the model, $r_1$ is the radius of the outer edge in AU (here $r_1\,=\,1.7$), and $\lambda_e$ is the angular separation of the Earth-to-pulsar line-of-sight from the Sun in the ecliptic plane. This contribution is added to the DM calculated by integrating through the WSA-ENLIL model for every pulsar LoS.

It should be noted that while WSA-ENLIL has been well validated, its ability to reproduce the actual density observed at Earth is far from perfect. The predicted density at Earth generally has much smoother time behavior than the actual observations demonstrate, although for the purposes of correcting DMs integrated through a large column of SW, smoothing over small-scale density fluctuations should not affect the results. \citet{2015SpWea..13..316J} found that the mean square error of the WSA-ENLIL density time series predicted for Earth, averaged to 1-hour time resolution, was around 20 cm$^{-6}$ (since this is a squared error, it is biassed upwards by periods of high SW density). This reflects the fact that predicting conditions throughout the SW with present resources is an intrinsically difficult problem.

\begin{figure*}
    \centering
    \includegraphics[width=\textwidth]{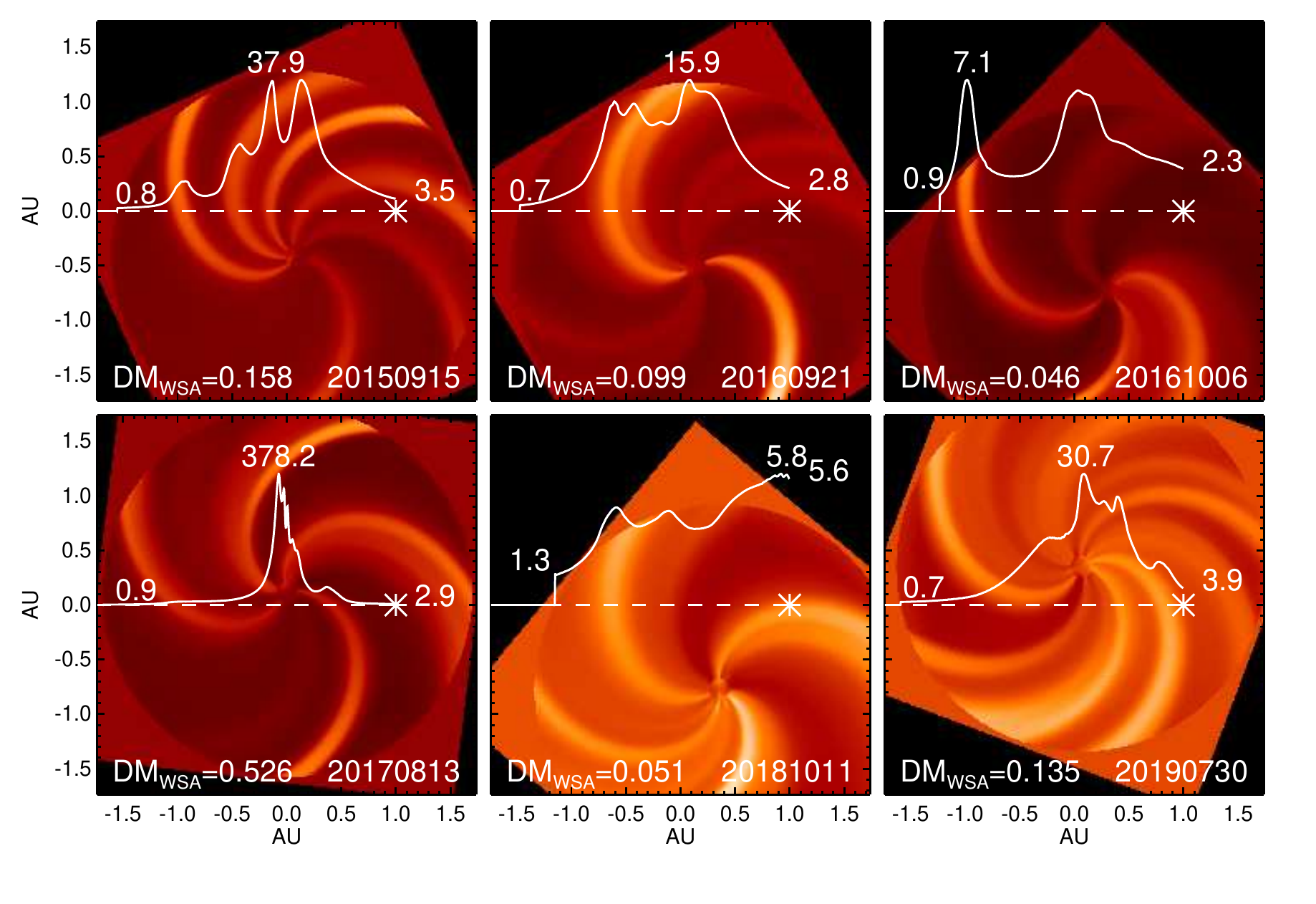}
    \caption{Illustrations of density profile measurement using \textsc{wsa}, for 6 epochs of B0950+08 at varying elongations. The circular part of each sub-image shows the density structure in the SW during a pulsar observation, divided by the azimuthally-averaged radial density profile in order to emphasize structure and de-emphasize radial dependence. The center of the circular region in each panel is the location of the Sun, the white asterisk towards the right marks the position of Earth, and the dotted line is the LoS to the pulsar (located to the left in each panel). The solid line is the density profile along the LoS, linearly scaled to the maximum density along the LoS (denoted above the peak). The SW density at Earth, and the density at the edge of the \textsc{wsa} region farthest from Earth along the LoS are also marked. The date is provided in the lower right of each panel, and the calculated \textsc{wsa} contribution to the DM is given in the bottom left.}
    \label{fig:1}
\end{figure*}

\vspace{-0.5cm}
\section{Data Reduction and Analysis} \label{sec:DRaA}
\subsection{Pulsar Timing and DM measurement}\label{sec:DRaA1}
For our current study, we analyze the data obtained at the three higher frequency tunings, as mentioned in section \ref{sec:observe}. We do not use the lowest frequency tuning of 35.1 MHz since it generally has poorer signal-to-noise (S/N) due to the reduced sensitivity of the LWA towards lower frequencies and greater contamination from radio frequency interference (RFI). Since the average shape of the pulsar profile varies with frequency, which can affect the accurate measurement of DM \cite[]{2007MNRAS.377..677A}, we treat each of these three tunings at 49.8 MHz, 64.5 MHz, and 79.2 MHz independently, and combine the reduced data at a later stage to obtain the best possible estimate of DM variation over time. First, we obtain a separate average pulse profile at each of the center frequencies. For this, we average observational data at one or more epochs, depending on the strength of the pulsar. First we apply a median ``RFI-masking" routine using PSRCHIVE, which removes data points higher than six times the median value for a smoothing window size of 13 channels in frequency. This is followed by visually inspecting each masked file for any possible low-level RFI which may be seen. These are then edited using the {\tt psrzap} routine in PSRCHIVE. The average profiles are then created by averaging in time and frequency to one sub-integration and one channel, followed by smoothing, using tasks {\tt pam} and {\tt psrsmooth} from PSRCHIVE. Also, in generating the average profile we only use those epochs when the pulsar was at least $40^{\circ}$ away from the sun. This was done to avoid the larger dispersion of the pulses due to the stronger contribution of the SW at smaller separations. These profiles were then modeled as Gaussian components, and all were aligned in phase with respect to the pulse profile at 79.2 MHz to get the correct pulse time-of-arrival (TOAs) for each independent tuning. These three final Gaussian profiles are treated as template pulse profiles for each tuning.

To calculate the DM from the entire data, we first apply a RFI excision scheme similar to one applied for creating the template. Since the archived data is already DM-corrected via coherent dedispersion using the ephemeris derived from LWA pulsar observations as stated in \citet{2015ApJ...808..156S}, we further reduce this cleaned data to one sub-integration and two or eight frequency channels based on source brightness by averaging in time and frequency, using the PSRCHIVE routine {\tt pam}. This is done independently for each tuning. We obtain the TOAs across the band, independently for each tuning using their respective template profiles via the PSRCHIVE task {\tt pat}. These TOAs are calculated via least-square fitting in the Fourier domain \cite[]{1992PTRSL.341..117T} by cross-correlation with the template profiles. Due to the small overlap in frequencies between the tunings, we generate four duplicate TOAs (for eight channels per tuning). This redundancy is removed, taking into account the roll-off of frequency filters towards the band edges. The rest of the 20 frequency-resolved TOAs, with $\sim$2.45 MHz bandwidth each, from all three tunings, are combined into a single file. For the case of data reduced to two channels per tuning (which is the case for PSR B1257+12), we obtain 6 independent ToAs, since there is no frequency overlap. We now use TEMPO to fit for the pulsar's spin and astrometric parameters using ephemerides as stated in \citet{2015ApJ...808..156S}, on all the obtained ToAs, to see if there is any significant variation in those parameters. The updated files are then used to get a measurement of DM variation (DMx) over time. This measures the change in DM with respect to the fiducial value as given in Table \ref{tab:1}, by fitting the TOAs for a time span over multiple epochs. The TOAs are fitted iteratively, cleaning the farthest outliers in each step until a $\chi^2$ value of $\sim 1-2$ is reached in the fitted residuals. This is done by visually inspecting TEMPO residual plots and removing the farthest outliers, and then refitting the remaining ToAs. The DMx values obtained at each epoch and their measurement errors are then stored separately for further analysis.

\subsection{Calculation of SW contributions}\label{sec:DRaA2}
For each pulsar, at each epoch observed, we calculate the angular separation from the Sun. This angle is supplied to the analytical models presented in Equations \ref{eq:3}, \ref{eq:4}, \& \ref{eq:5} in order to calculate the respective SW DM contributions in pc cm$^{-3}$. The distance between the Sun and Earth is also needed for the spherical wind models, where the DM contribution is calculated by summing, along the appropriate line of sight, the contributions between successive spherical shells centered on the Sun, using a resolution of 0.01 AU in the radial direction. Given the number density values presented in section \ref{sw}, along with Equation \ref{eq:3}, contributions from the spherical SW models T1 and T2 are computed. Similarly, contributions from \textsc{fast} and \textsc{slow} SW models are computed using Equations \ref{eq:4} and \ref{eq:5} respectively.

In the case of \textsc{wsa}, the model provides a density profile along the line of sight as a cut through the ecliptic. Figure \ref{fig:1} shows examples of the density profiles along the LoS to B0950+08 at six epochs representing different angular separations. The patterns across the ecliptic plane reveal the presence of the higher- and lower-density streams discussed earlier, in the form of streams with a spiral shape induced by solar rotation. In cases where the LoS to the pulsar is close to the Sun, the SW DM is dominated by the density closest to the Sun (sharp peaks towards the center), while at larger separations, dense streams produce the largest contributions. We do not evaluate the \textsc{wsa} model for B1257+12, since it has high ecliptic latitude and we are uncertain how far out of the ecliptic plane the model can be regarded as valid.

\begin{figure*}
    \includegraphics[width=\textwidth,height=0.75\paperheight]{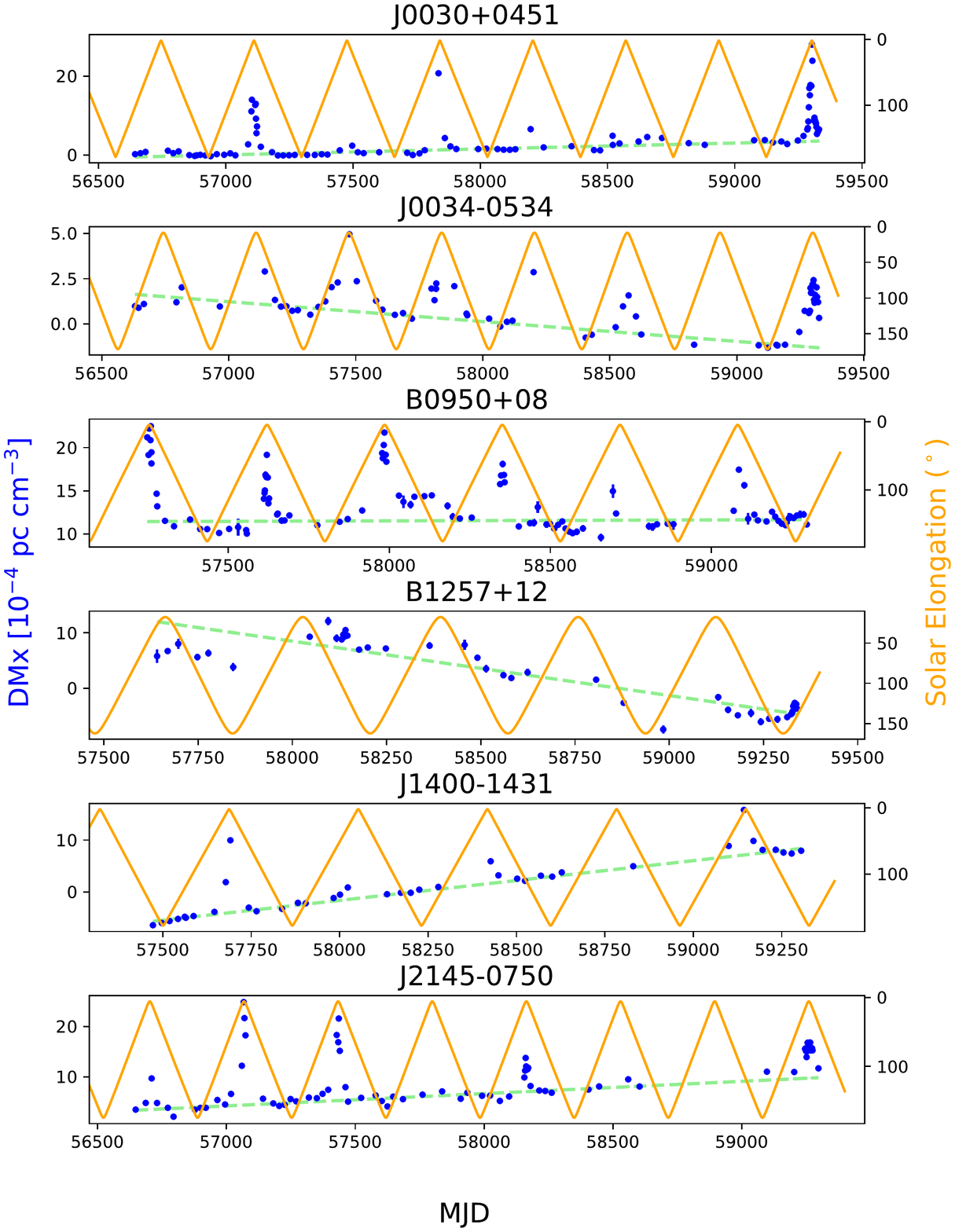}
    \caption{DM variation (DMx) timeseries for our pulsar sample. The blue points show the measured variations along with error bars measured using pulsar timing. The error bars are generally much smaller than the DMx values and so are not visible on the plot. The orange curve shows the angular separation between the pulsar and the Sun with the corresponding scale shown on the right axis. Each panel corresponds to a different pulsar. The dashed light green line shows the linear trend fitted as the IISM contribution.}
    \label{fig:3}
\end{figure*}

\subsection{Modeling the IISM contribution}\label{sec:DRaA3}

In general both the SW and the IISM are contributing to the DMx contribution in any given measurement, and the two contributions need to be separated if we are to evaluate the effectiveness of SW models. We use the fact that the SW contributions to the DM are expected to be significant at small solar elongation angles, whereas at larger separations one expects DM variability to be dominated by the IISM. Our inferred SW contributions for different models were found to be about an order of magnitude smaller than the measured DMx values at angular separations $>60\degr$ from the Sun\footnote{Note that we do not calculate contributions from the \textsc{wsa} model for separations beyond $60\degr$, since the outer radius of 1.7 AU in the model means that not much of the wind is being sampled.}. This suggests that even though there remains a scatter in the measured DMx values with solar elongation for some pulsars, those measurements are not dominated by the SW. Hence, as a starting point to model the contribution to DMs from the IISM, we choose a binning of our DMx values beyond $60\degr$ separation and apply a polynomial fit to estimate the expected IISM contribution at all separations. A linear fit was found to be sufficient to model these residuals, based on an F-test performed on the $\chi^{2}$ values of the fits for polynomials of order 1, 2 and 3. Applying the same procedure after binning the data at $>30\degr$ separation in steps of $5\degr$ also gave similar values for the final residual rms. Other smaller separations were not tried because of possible contamination from the SW at smaller separations.

\begin{figure*}
    \centering
    \includegraphics[scale=0.55]{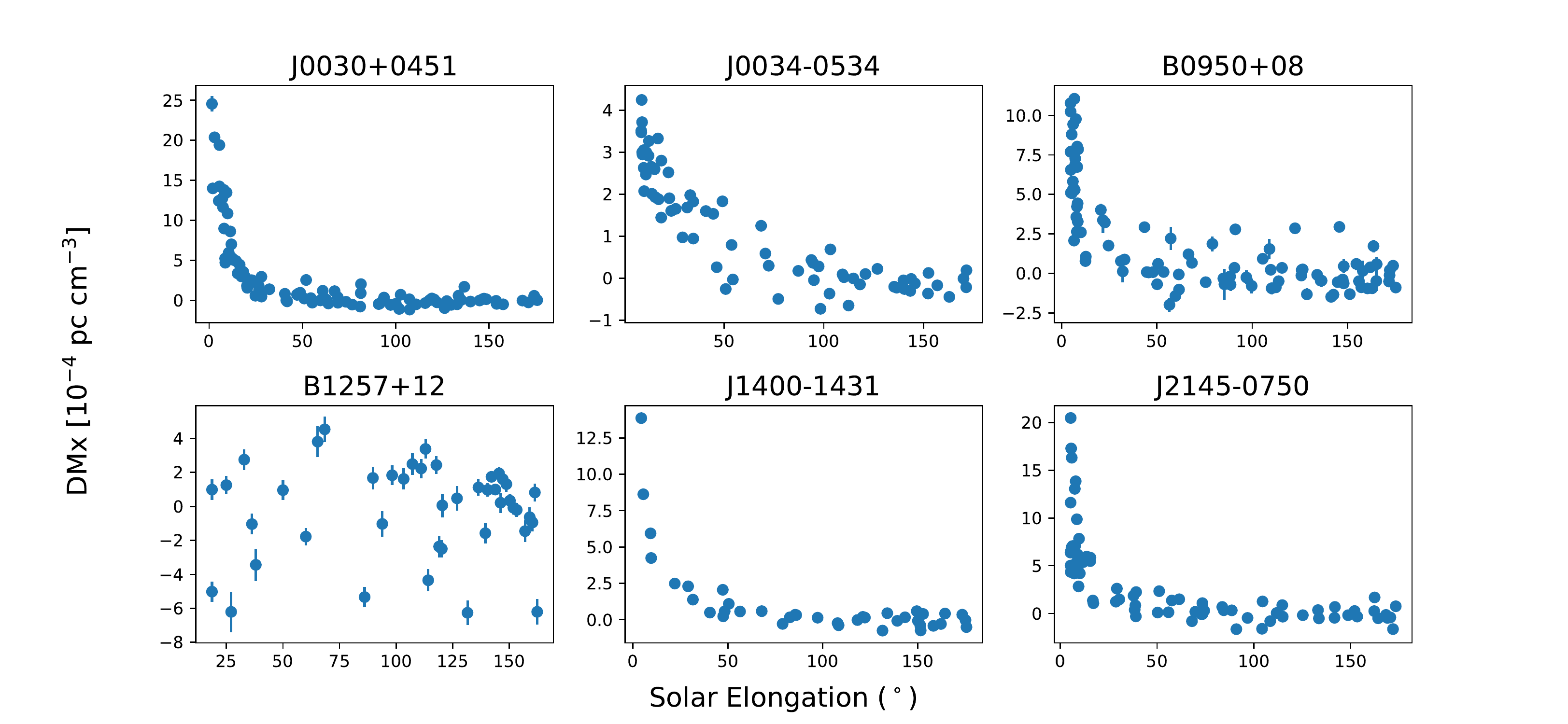}
    \caption{The plot of measured DMx with errorbars as a function of separation between the pulsar and the Sun. Each panel corresponds to a different pulsar. The fitted linear trend of the expected IISM contribution, as shown by the light-green dashed line in Figure \ref{fig:3}, has been subtracted from the DMx values in each case to show the effect of SW contribution on DMx with solar elongation angle.}
    \label{fig:4}
\end{figure*}

The resulting linear fits to the IISM contribution are shown as green lines in Figure \ref{fig:3}, where the DMx values are plotted as a function of date for all 6 pulsars. Figure \ref{fig:4} shows the resulting dependence of DMx on solar elongation angle after the fitted IISM contributions have been subtracted: for all of the pulsars except B1257+12, the remaining DMx contributions show a pronounced dependence on solar elongation that starts to flatten out around zero beyond about $25-30$ degree separation. Nevertheless, the data for B1257+12 show that this is not always the case. This may suggest that the IISM and SW behavior are entangled at some level, and perhaps a more complex approach is required to separate cleanly the two contributions. A similar conclusion was also drawn by \citet{Tiburzi2019}.

To minimize the possibility of entanglement in our IISM modelling, we also tried piecewise-fitting of the data on annual timescales, achieved by binning the data for separations larger than $60\degr$, grouped between consecutive yearly near-Sun transits of the pulsar. No statistically significant improvements were seen based on an F-test performed on respective $\chi^2$ values from continuous and piecewise fitting for the case of a linear polynomial, except for PSR J1400-1431. However, even in that case the final residual rms for this pulsar did not improve as a result of the piecewise fitting. A similar analysis was not performed using higher-order polynomials due to having insufficient points in a given bin to constrain the fitting parameters. Based on these results, we do not find any value in fitting more free parameters to the IISM variation than are needed for the linear polynomial fit in time using DMx values measured beyond $60\degr$ separation. 

\begin{table*}
    \begin{center}
    \tabcolsep=0.01cm
    \begin{tabular}{cccccc}
    \\[-1.8ex]\hline  
    \hline \\[-1.8ex]
      &J0030+0451&\hspace{0.5cm}J0034-0534&\hspace{0.5cm}B0950+08&\hspace{0.5cm}J1400-1431&J\hspace{0.5cm}J2145-0750\\
    \hline
    T1&13.2&51.2&0&2.2&21.7\\
    \hline
    T2&18.1&11.7&3&11.7&30\\
    \hline
    \textsc{slow}&21.1&71.2&0.7&4.6&28.5\\
    \hline
    \textsc{fast}&42.5&44.4&7.8&26.5&45.5\\
    \hline   
    \textsc{wsa}&14.6&3.6&1.7&-1.5&16.8\\
    \hline
    \end{tabular}
    \end{center}
    \caption{This table shows the percentage change in calculated RMS values between cut-off angular separations of 40 and 15 degrees, with respect to the value at 40 degrees, for different pulsars and SW models.}
    \label{tab:3}
\end{table*}

\begin{table*}
    \begin{center}
    \tabcolsep=0.01cm
    \begin{tabular}{ccccccc}
    \\[-1.8ex]\hline  
    \hline \\[-1.8ex]
      &\hspace{-5cm}J0030+0451&\hspace{0.5cm}J0034-0534&\hspace{0.5cm}B0950+08&\hspace{0.5cm}B1257+12&\hspace{0.5cm}J1400-1431&J\hspace{0.5cm}J2145-0750\\
    \hline
    \begin{equation*}
    \frac{dDM}{dt}
    \hspace{0.2cm}(10^{-5} \frac{pc}{cm^{3} yr^{1}})    
    \end{equation*}&\hspace{-5cm}6.5$\pm$1.2&-5.0$\pm$1.1&0.5$\pm$1.9&-31.5$\pm$5.3&28.0$\pm$1.2&10.5$\pm$1.6\\\\
    \hline
    \begin{equation*}
    M(eDM) \hspace{0.2cm}(10^{-5} \frac{pc}{cm^{3}})    
    \end{equation*}&\hspace{-5cm}3.0&0.65&2.67&5.98&0.94&2.53\\
    \hline
    \end{tabular}
    \end{center}
    \caption{This table shows the average time derivative of DM after subtracting solar wind amplitudes and the median DM uncertainty of our data. The time derivative is calculated after binning the data SW model subtracted data for $>10 \degr$ elongation angle.}
    \label{tab:4}
\end{table*}

\begin{figure*}
    %\vspace{-3cm}
    \centering
    \includegraphics[width=\textwidth,height=0.7\paperheight]{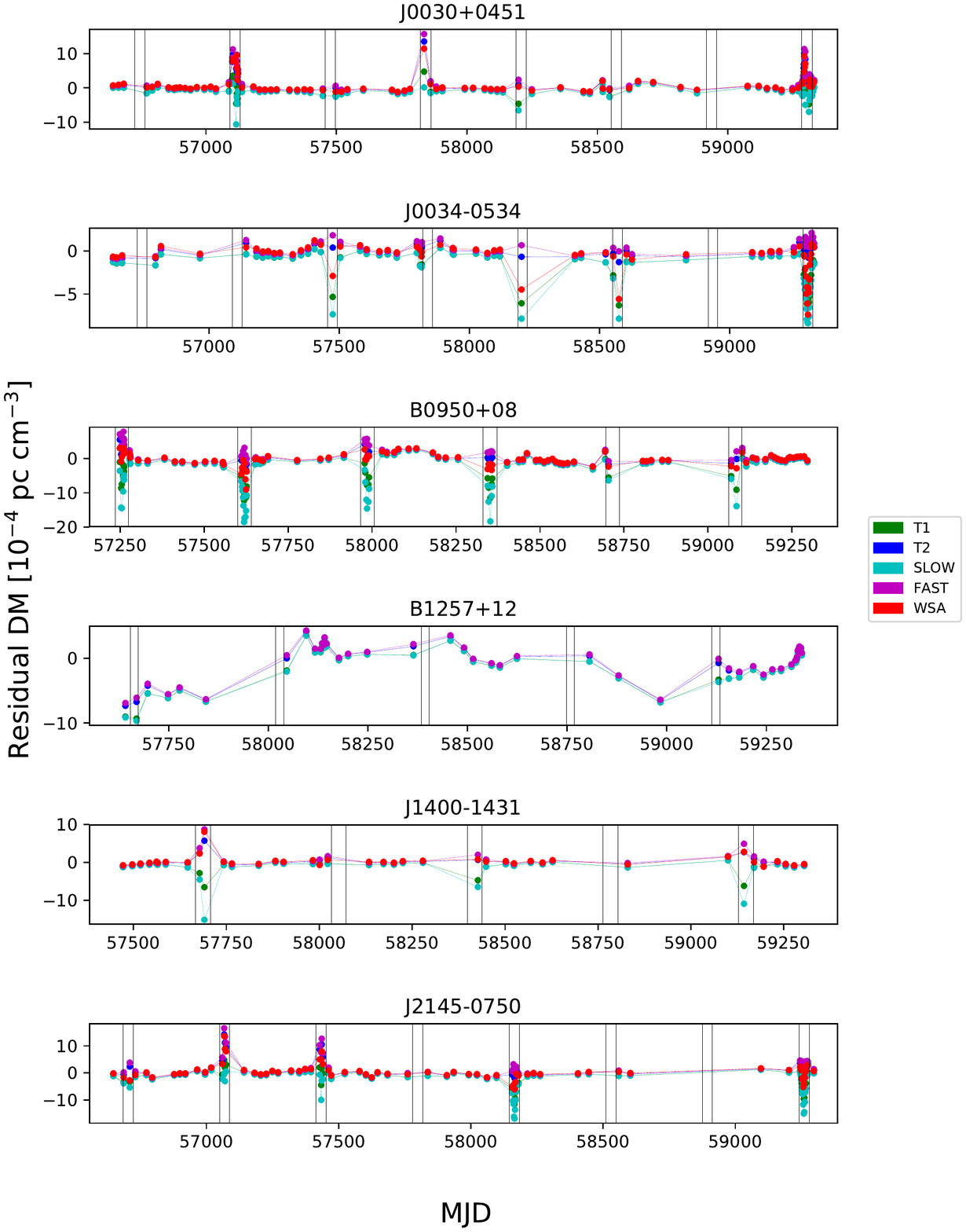}
    \caption{The final DM residuals after SW and IISM contributions are subtracted from the measured DMx. The different colors in the legend show the different SW models used to compute the residuals. The vertical black lines in the subplots mark the periods when the pulsar is at an angular separation of $<20\degr$ from the Sun. For PSR B1257+12 there is no \textsc{wsa} plot since we did not calculate the \textsc{wsa} values in this case due to the higher ecliptic latitude of the pulsar. Residuals are plotted only for $>4\degr$ separation because of large deviations below that. }
    \label{fig:5}
\end{figure*}

\begin{figure*}
    \centering
    \includegraphics[scale=0.55]{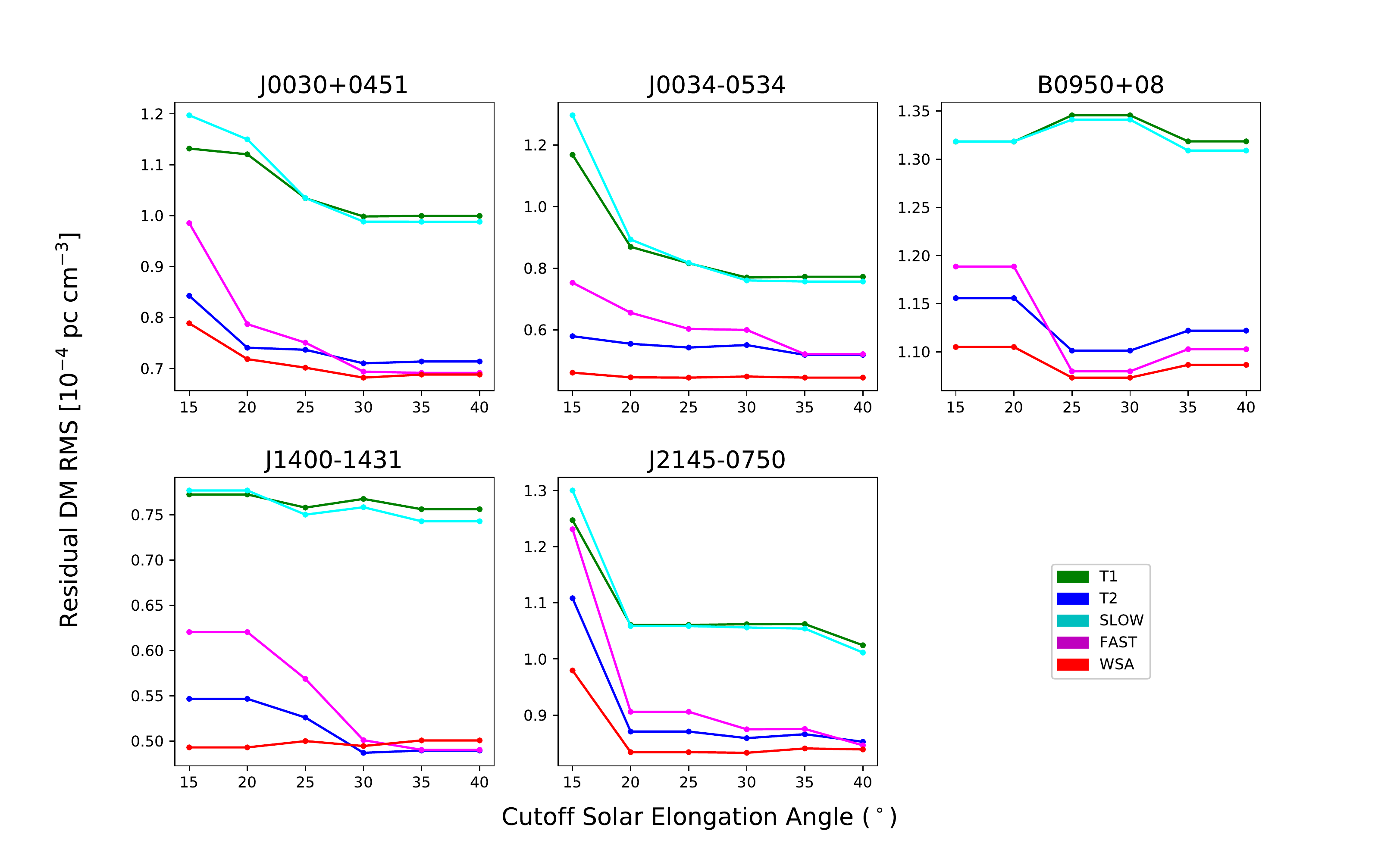}
    \caption{The calculated RMS of the DM residuals as a function of the cutoff solar elongation angle at 5\degr\ increments. The different colors show the different SW models used to calculate the residuals, as marked, and the panels correspond to different pulsars.}
    \label{fig:8}
\end{figure*}

\begin{figure*}
    \includegraphics[scale=0.55]{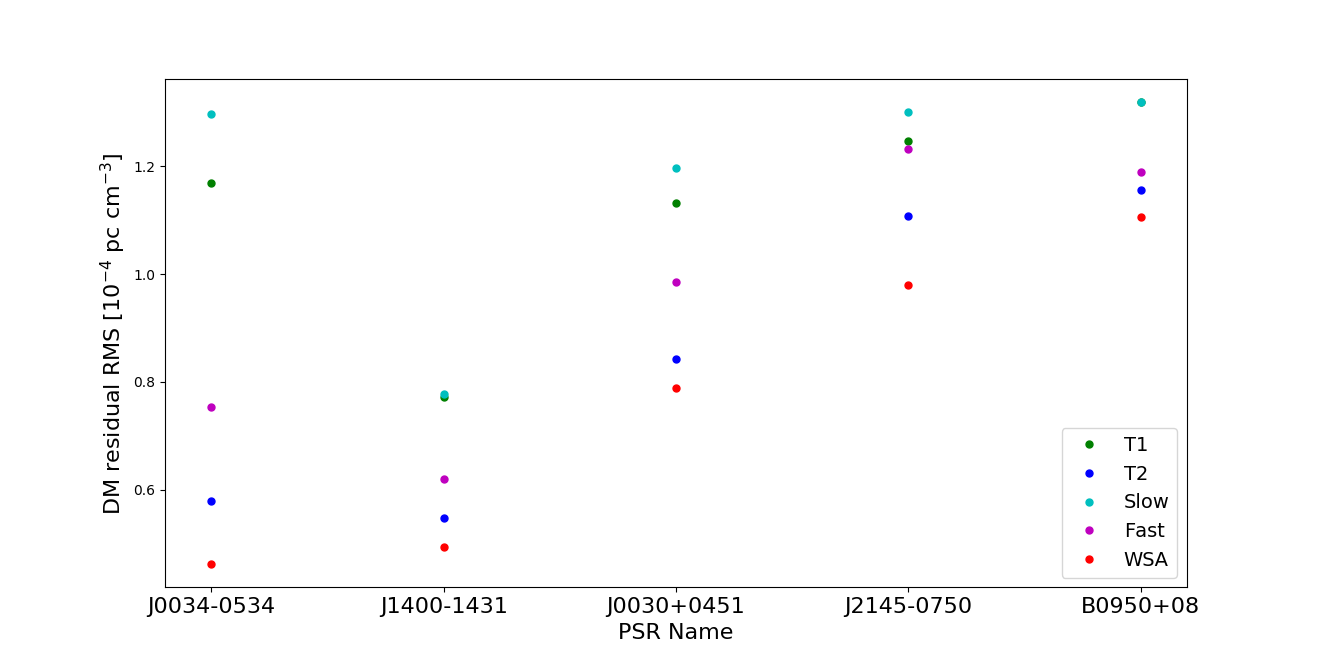}
    \caption{The RMS of the residual DM beyond 15\degr\ separation for each pulsar, with the results for the different SW models given by different colors in the plot legend. On the horizontal axis, the pulsar names are plotted in order of increasing spin period from left to right, based on the values given in Table \ref{tab:1}}
    \label{fig:6}
\end{figure*}

\vspace{-0.3cm}
\section{Results}\label{results}
We analyze the DM variation with time for six pulsars over 7 years, including five MSPs and a slow pulsar, and investigate the effectiveness of a range of SW models to account for the measured variations.

\subsection{DM variation over time} \label{results1}
Figure \ref{fig:3} shows the variation of measured DM for the pulsar sample over time along with the corresponding solar elongation angle. The variations have been measured over $5-7$ years, depending on available data for different pulsars. Each of the pulsars shows evidence for a linear trend with time in DM at large elongations that can be attributed to variation in the IISM along the LoS to the pulsar. In particular, J1400-1431 shows a relatively sharp increase with time. The rate of change of the IISM contribution is not simply related to the total column to the pulsar, since J1400-1431 has a relatively low DM,  similar to that of B0950+08 and J0030+0451, which do not show similar rates of change.

The behavior of B1257+12 appears somewhat anomalous compared to the other 5 pulsars. While there is a general linear trend of decreasing DMx with time, it does not show the same striking dependence on solar elongation that the other pulsars share (Figure \ref{fig:4} shows the dependence of pulsar DM after correction for IISM variation). Partly this is because its ecliptic latitude (17.6\degr) is significantly higher than the other sources, so its LoS never passes as close to the Sun, but there does seem to be little effect of solar distance in its data, and it exhibits a much larger scatter in measurements for elongations $>30$\degr\ than the other pulsars. PSR B1257+12 is a complex system: the pulsar is in orbit with a planetary system \citep[e.g.,][]{1994Sci...264..538W}, and it may be that some of the orbital variations produce timing offsets that are not adequately fit by the orbital parameters in the timing model, and instead are absorbed in our analysis as spurious DM variations.

\subsection{Comparison of SW models} \label{results2}
Figure \ref{fig:5} shows the DM residuals (DMx-SW-IISM, hereafter referred to as the residual) for the six pulsars with respect to different SW models, as a function of time. Regions of angular separations from the Sun below $20\degr$ are marked using black vertical lines.  At larger angular separations, the residuals are close to zero. When the pulsar approaches the Sun, residuals are large, and no SW model provides a suitable correction. RMS values for the residuals were computed by binning them in the $15\degr-40\degr$ elongation range in steps of $5$ degrees, as shown in Figure \ref{fig:8}. Table \ref{tab:3} reports the changes in RMS between $40$\degr\ and $15$\degr\ separation angle, with respect to the value at $40$\degr\ separation. This was found to be $30 \%$ or more for SW models other than \textsc{wsa}, with deviations in some cases as high as $70 \%$. For the case of the \textsc{wsa} SW model, the change was $\sim 15 \%$ or less for all pulsars. A similar comparison between $40$\degr\ and $10$\degr\ angular separation showed significantly higher RMS values, with up to twice larger values in some cases. Figure \ref{fig:6} shows the residual RMS for each pulsar for each SW model. The lowest residual RMS are obtained in the case of the \textsc{wsa} model for each pulsar. A trend in residual RMS is seen, in that the RMS seems to increase with the pulsar spin period. A similar plot against ecliptic latitude did not show any trend in the calculated RMS values. Given the small number of pulsars in our sample it is difficult to claim if this trend is due to some causal relationship or related to some systematic in the data. For PSR B1257+12, the computed RMS were found to be significantly higher than the other pulsars, of order of $10^{-4}$ pc cm$^{-3}$.

\subsection{DM derivatives}\label{results3}
We quantify the overall DM trends by fitting a linear gradient to the data, using a least-square fitting routine, similar to \citet{2004MNRAS.353.1311H}. However, since our sample data is solar-wind dominated at small solar elongations, we apply the gradient fitting after subtracting the calculated SW contribution to the DM, to avoid bias. As we work with five different models of SW, it results in five different values of these derivatives. Nevertheless, we see that these values, after different SW amplitude corrections, converge for data binned at $>10 \degr$ elongation angle. This is consistent with our previous results, where we see large amplitudes of DM residuals for small solar elongations. Hence, we report the amplitude of our DM derivatives as the mean of these five values (four in the case of B1257+12), where the error is calculated by adding the individual errors in quadrature. The corresponding values are reported in Table \ref{tab:4}. To find the correlation between the DM and the amplitudes of its time derivatives, we fit a power law to this data. We get an exponent of $0.29 \pm 0.65$. Since this is consistent with the square root dependence in \citet{2004MNRAS.353.1311H}, we also fit for the amplitude explicitly assuming a square root dependence, yielding |$\frac{dDM}{dt}$|=0.00005$\sqrt{DM}$ for this small sample of pulsars, with negligible errorbars on the amplitude. This is a factor of 4 smaller than the \citet{2004MNRAS.353.1311H} estimate.

\section{Discussions} \label{sec:discuss}
For the sample of six pulsars examined in this study, each shows a significant DM variation over $5-7$ years. Measurement errors on DMx are an order of magnitude or lower than the corresponding DMx values for most observations, other than for PSR B1257+12, which is relatively weaker in our frequency bands. There is a clear dependence of the measured DM variations on the solar elongation angle for B0950+08, J0030+0451, J0034-0534, J1400-1431, and J2145-0750 marked by the sharp rise in values at separations less than $\sim 25-30$ degrees, as shown in Figure \ref{fig:4}. 

For J1400-1431, we see the SW effect superimposed on a sharp upward trend of DMx, in time, as shown in Figure \ref{fig:3}. The robust linear increase in time for this pulsar is likely due to fluctuations in the local interstellar medium, as discussed below, and merits further investigation. It is worth pointing out that B0950+08, J0030+0451, and J1400-1431 are the lowest-DM pulsars in the sample with comparable distance estimates as reported in the ATNF\footnote{\url{https://www.atnf.csiro.au/research/pulsar/psrcat/}} \cite[]{atnf2005AJ....129.1993M} pulsar catalog. Based on the proper motion values (in mas yr$^{-1}$),  J1400-1431 is the fastest moving pulsar, about twice faster than B0950+08, which is closer than the former. However, B0950+08 does not show a trend in DMx similar to that of J1400-1431, even if we were to consider data at separations of more than $40$, $50$, and $60$ degrees from the Sun. This suggests that the strong trend in DMx values for J1400-1431 cannot be explained as a simple increase in path length through the IISM due to the pulsar's proper motion, but rather suggests a role for spatial variations in the IISM.

B1257+12 does not show the same robust dependence of DMx on solar elongation angle seen in the other pulsars. Since its closest approach to the Sun is $\sim 17$ degrees and, as shown in Figure \ref{fig:3}, there is no clear correlation between its closest approach and corresponding peaks in the DMx values, it is possible that the DM variations are due to the combined effect of the SW and the complex orbit, which causes the scatter in DMx evident even at large separations.

 Figure \ref{fig:3} suggests some evidence for a change in the SW contribution over time due to the solar cycle: the height of the peaks in DMx at small elongations appear to change with time, although the sampling is insufficient for a strong conclusion. As an example, we measure DMx values of about $1.2$, $2.0$ and $1.7$ ($10^{-3}$ pc cm$^{-3}$) for J0030+0451 at $\sim 5$ degree elongation from the Sun, at different times, with similar errorbars, which are much smaller than the separation between these values. The same trend was seen even without any IISM component subtraction from the DMx data. In principle the \textsc{WSA} model handles solar cycle variations since it is driven by measurements of the actual solar magnetic field, but the other models do not: this would require modeling of SW contributions with variable amplitude \cite[see, e.g.,][]{you10.1111/j.1365-2966.2012.20688.x,Tiburzi2019}.\\
 The calculated values of our DM time derivatives are  comparable to those presented in \citet[]{2004MNRAS.353.1311H} and \citet{jones17} for PSR B1257+12 and J2145-0750, whereas for the other pulsars in common, the former has values which are two or more orders of magnitude larger than those presented here. A similar comparison shows that our values are in agreement with those reported in \citet{donner2020},  except for the case of B1257+12, which shows anomalous behavior. One important difference between the two cases is the observing frequency. While the former two are primarily at high frequencies (>300 MHz), the latter is below 200 MHz, closer to the LWA frequencies which are more sensitive to DM variations. This reinforces the argument presented in \citet{donner2020}. Similarly, comparison of correlation between the amplitude of the DM time derivative and the DM magnitude shows that the time derivative follows a square root dependence consistent with \citet[]{2004MNRAS.353.1311H} and \citet{donner2020}. However, the amplitude of the square root dependence found here is comparable only with the latter study, being an order of magnitude smaller than the former.

\begin{table*}
    %\begin{center}
    \centering
    \tabcolsep=0.1cm
    \begin{tabular}{cccccc}
    \\[-1.8ex]\hline 
    \hline \\[-1.8ex]
    &J0030+0451&\hspace{0.5cm}J0034-0534&\hspace{0.5cm}B0950+08&\hspace{0.5cm}J1400-1431&\hspace{0.5cm}J2145-0750\\
    \hline
    $RMS_{15}$ ($\mu $s)&0.167&0.098&0.234&0.104&0.207\\
    \hline
    $RMS_{40}$ ($\mu $s)&0.146&0.094&0.230&0.106&0.178\\
    \hline
    \end{tabular}
    %\end{center}
    \caption{Table of expected RMS of time delay at 1.4 GHz based on the residual DMs calculated at our frequencies. $RMS_{40}$ is for angular separation of more than $40$ degrees and $RMS_{15}$ is similarly for $15$ degrees separation.}
     \label{tab:2}
\end{table*}

\begin{figure*}
    %\epsscale{0.5}
    %\plotone{new_time_delay_new.eps}
    \includegraphics[width=\textwidth,height=0.65\paperheight]{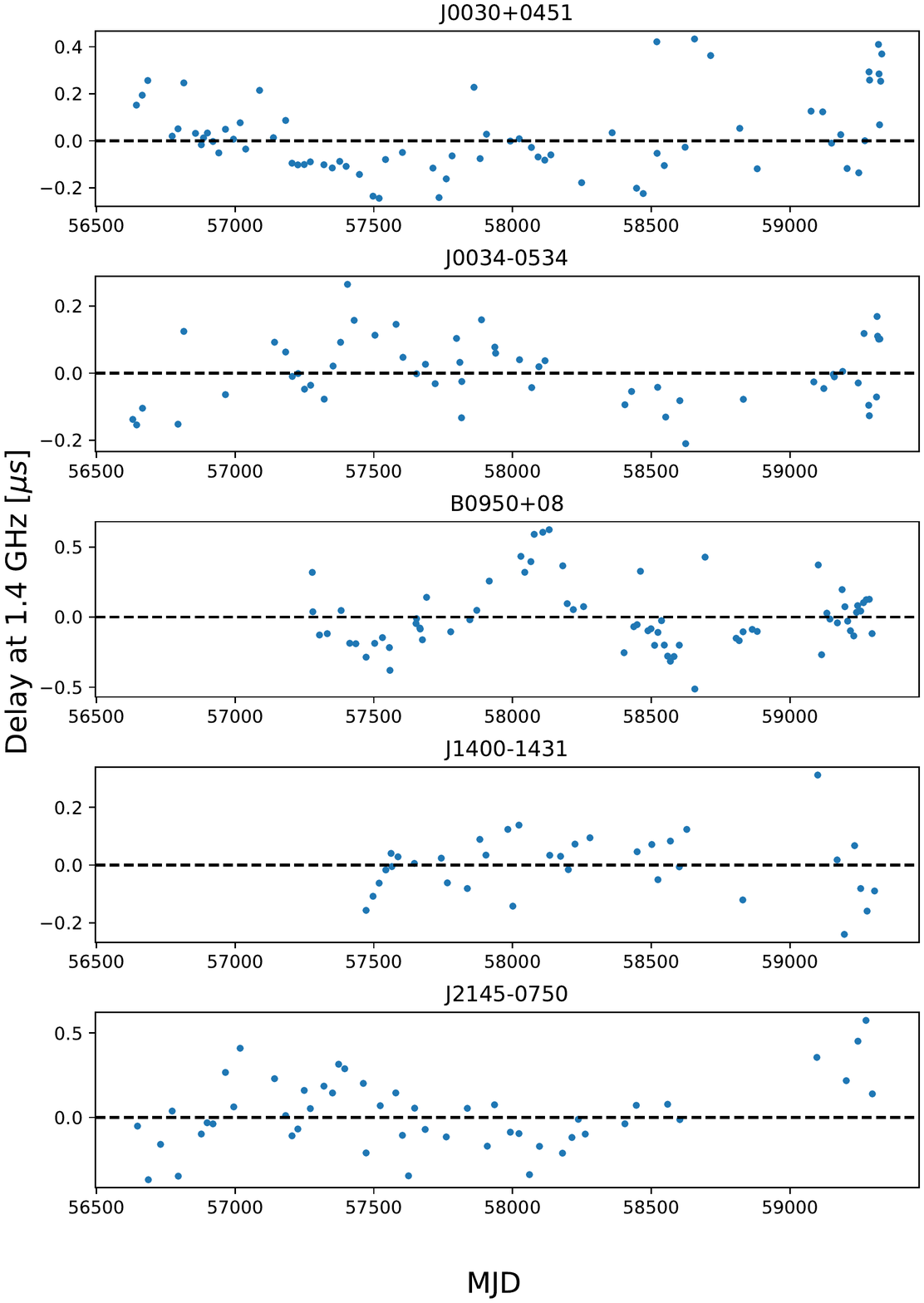}
    \caption{The calculated time delays at 1.4 GHz for angular separations of more than $15$ degrees solar elongation angle. The values are based on the corresponding DM residuals for  \textsc{wsa} SW model.}
    \label{fig:7}
\end{figure*}

\subsection{Efficacy of SW models and IISM modeling}\label{sec:discuss1}
The residual DMs shown in Figure \ref{fig:5} indicate that all SW models perform equally well at larger angular separations, but the situation changes at small elongations. The \textsc{slow} and \textsc{fast} wind models generally overestimate and underestimate the SW contributions, respectively, at smaller separations, creating larger DM residuals. Also, the spread in residuals (the measured RMS value) increases more steeply with decreasing separation for these models, as mentioned in section \ref{results2}, suggesting that the near-Sun application of these models is inadequate to capture fully the effects of the SW.

The spherical approximation models T1 and T2, where the former assumes the SW to be predominantly slow and the latter assumes it to be fast, perform slightly better than their respective \textsc{slow} and \textsc{fast} counterparts. The residuals, in this case, are smaller near the closest approach compared to the previous two. Moreover, the change in RMS with decreasing elongation for these two models is more gradual than the \textsc{slow} and \textsc{fast} cases. 

The \textsc{wsa} operational model is found to give the best representation of the SW contribution to DMs, as the residuals are consistently the lowest and remain effective at smaller angular separations. The change in RMS with elongation is also more gradual, remaining below $15 \%$ for all pulsars, whereas for other SW models, even though the change was $< 25 \%$ for most pulsars, in some instances, it was up to $50 \%$ or more. Hence, \textsc{wsa} keeps the residual RMS lower, irrespective of the pulsar. This is shown in Figure \ref{fig:8}, where the change in residual RMS with elongation for the \textsc{wsa} SW model is visibly smoother than for the other SW models.

As shown in Figure \ref{fig:6}, the measured RMS of the DM residuals are of the order of several $\times\ 10^{-5}$ pc cm$^{-3}$ for the MSP sample and slightly higher in the case of the slow pulsar B0950+08. The overall trend in the calculated RMS remains consistent across all pulsars, in decreasing order of \textsc{slow}, T1, \textsc{fast}, T2 and \textsc{wsa} SW models.

The appropriate modeling of the interstellar variations is also necessary in order to minimize DM residuals, especially at large separations where the SW effects are less significant. As discussed in section \ref{sec:DRaA3}, a few different approaches to this problem were investigated, but it was found that  modeling the IISM behavior as a linear polynomial in time is sufficient at the measurement level achieved in these data, at least over the 7-year timescale considered here. Even if this does not fully disentangle the SW and IISM effects, this should account for the slow variations due to the small change in the LoS to the pulsar, caused by the proper motion of a pulsar through the spatially-varying interstellar medium.

\subsection{Comparison with other low frequency observations}\label{sec:discuss3}
For the common pulsars in our sample, we see similar trends in DM variations over long time scales as reported in \citep[see;][]{donner2020,Tiburzi2021, kkumar}, where the former two are studies conducted below 200 MHz and the latter reports on observations above 400 MHz. While these are in agreement on the overall DM variation for these pulsars for the given span of time, the median uncertainties on our measured DM values, as reported in Table \ref{tab:4}, are a factor 5-10 better than those in these studies, except for the case of B1257+12, where we get comparable values. Comparison of the expected RMS of time delay at 1400 MHz, as reported in Table \ref{tab:3}, for the \textsc{wsa} model, with those shown in \citet{Tiburzi2021} Figure 5, shows that the \textsc{wsa} correction are better by a factor of 2 or more, depending on the pulsar, where \cite{Tiburzi2021} uses a constant and variable amplitude spherical solar wind model. This suggests that corrections provided by models informed by observations, like the WSA, provided improved corrections to SW effects.

\subsection{Implications for PTAs}\label{sec:discuss2}
Inhomogeneity in the intervening medium can produce refractive effects and scattering of pulsar signals, resulting in different frequencies traveling slightly different path through the medium. When the medium has significant fluctuations in free electron density on small spatial scales, the effective DM contribution seen by different frequencies can vary. Since the IISM is inhomogeneous, this is a direction-dependent effect. In order to assess the impact of our results on timing arrays that typically operate at frequencies higher than does LWA, we assume a frequency-independent value of DM \cite[see, e.g.,][]{cordes2016ApJ...817...16C,lam2020ApJ...892...89L} and use Equation \ref{eq:1} to calculate the expected timing delay at 1.4 GHz, which is the commonly used frequency for PTA experiments. Figure \ref{fig:7} shows the calculated delays for more then $15$ degree separation, based on the DM residuals reported for the \textsc{wsa} model in Figure \ref{fig:5}. This is done by binning the DM residuals for $>15$ degree separation, for each pulsar after applying the \textsc{wsa} model and IISM contributions, and then calculating the expected delay for those values at 1.4 GHz using equation \ref{eq:1}. The resulting RMS values of the timing residuals binned at $15$ and $40$ degree separations are reported in Table \ref{tab:2}. We see that values are $< 300$ ns in all cases, and in some cases are as small as $100$ ns even at $15$ degrees, as desired for PTA experiments. The expected RMS of timing residuals will be higher in the case of the other SW models discussed here. Nevertheless, the RMS of DM residuals achieved for our sample is of the order of $10^{-5}$ pc cm$^{-3}$ for PTA-class pulsars, which is significantly better than the reported median DM uncertainty of the order of $5 \times 10^{-4}$ pc cm$^{-3}$ for PTA experiments \cite[see, e.g.,][]{2021ApJS..252....5A}. Hence, low-frequency observation of those pulsars could improve the noise limits in timing experiments by reducing the timing noise due to DM uncertainties. There are two possible scenarios, the first being a simultaneous observation of the pulsar at low frequencies and using the measured DM to correct the TOAs in the high-frequency data. The second is regular monitoring of the pulsar at low frequencies to model the slow trends due to the IISM, combined with high cadence observations (possibly same day) during the closest approach of the pulsar LoS to the Sun to obtain corrections due to the SW. As the SW is highly variable, low-frequency observations close to the high-frequency PTA observations are necessary to avoid introducing excess noise in the data. Some improvement could also be obtained by applying the \textsc{wsa} model rather than the default model used in PTA data analysis. Also, as we move into an observing cycle of increasing solar activity, using a model which is not stationary in time would be more appropriate to mitigate the SW effects, since it can capture the variable nature of the SW. There are two different approaches which could be employed to this effect: 1) using an existing SW model like the \textsc{wsa}, which is driven by parameters from independent cotemporaneous observations of the Sun, or 2) to perform very high cadence observations of a pulsar at low radio frequency and using these observations to perform an amplitude scaling of the stationary models of the SW such as T1, T2 etc., which has already been suggested in some previous studies. However, any such amplitude modeling would need to be tested against other measurements of SW, so as to avoid any biases which may be introduced due to pulsars in the sample data.

\vspace{0.5cm}
\section{Summary and Conclusions}
We have studied SW effects and IISM variation in a sample of six pulsars using $5-7$ years of available data. Five different SW models have been investigated, providing different levels of precision for DM measurements. The commonly used spherical SW models perform better than assumptions of slow or fast approximations of the SW. The measured DM residuals are comparatively smaller, in this case, up to solar elongation angle of $ \sim 25-30$ degrees. However, there is a swift change in RMS values at closer separations due to insufficient correction for the SW. As the simplest spherical models have a time-constant amplitude of SW correction, they also fail to account for the change in SW activity during a solar cycle on longer time scales, but a time-dependent version of the model can be applied, as shown by \citet{Tiburzi2019}. Nevertheless, the current study shows that the spherical model does not work equally well for all pulsars, as some pulsars show a more significant change in residual RMS as the pulsar approaches close to the Sun.

\textsc{wsa} is the other SW model investigated in this article, applied to operational space weather forecasting by using observations of solar magnetic fields to drive a time-dependent model of the SW, as well as to catch CME transients directed towards the Earth. Implementation of the \textsc{wsa} model for measuring the SW effect in low-frequency timing data provides the smallest DM residuals. At the same time, it is equally effective for all pulsars in our sample, with a change in residual RMS below $15 \%$ in all cases between $40$ and $15$ degree angular separations. Hence, this model remains effective even at a closer approach to the Sun. In order to constrain the RMS of DM residuals given by these cutoff angular separations, we need to avoid observation windows of $30$ and $80$ days for $15$ and $40$ degree cutoff angular separations between the pulsar LoS and the Sun. This would amount to losing $\sim$ $9 \%$ and $22 \%$ of the total observing time, respectively.

These SW models can be used to some extent down to $10$ degree separation, but with significantly higher residuals. None of the models work well below $10$ degrees. However, since high precision pulsar timing experiments generally tend to avoid an observation window ($\sim 10-15$ days) near the solar conjunction of the pulsar, it is safe to assume that corrections at those separations would not be required. 

In this study we used archival \textsc{wsa} data from the daily operational runs provided by NOAA in order to test the value of more sophisticated SW models for correcting pulsar DMs. This data source has the advantage that the complex 3D MHD models have already been run for essentially every day for which we have pulsar data, but the disadvantage that (in order to reduce file sizes) only 2D data in the ecliptic plane can be used and this is obviously not satisfactory for wider application. However, suitable 3-D models are available: NASA's Community Coordinated Modelling Center provides runs-on-demand for a number of models, including WSA\footnote{\url{https://ccmc.gsfc.nasa.gov/models/index.php}}. Other approaches may also prove fruitful, such as the SW model derived from observations of interplanetary scintillation \citep{2011JASTP..73.1214J}. We are in the process of exploring approaches to provide high-quality 3D SW models suitable for routine estimation of DM contributions for pulsar timing applications.

Even though we find that these SW models do not consistently reach the $100$ $ns$ timing accuracy desired by PTA experiments, using these SW models and low-frequency pulsar observation, we were able to achieve an RMS on DM residuals of the order of several $\times\ 10^{-5}$ pc cm$^{-3}$, which is better than existing values for Pulsar Timing Arrays. Hence, using low-frequency observations could be helpful to improve the noise limits in PTA experiments by minimizing the effect of DM variations and allowing for the inclusion of observations at smaller solar elongation angles. Low-frequency pulsar observations can also be used to test the efficacy of other space weather models. We also found that the variations in IISM contribution were slower than suggested by \citet{2004MNRAS.353.1311H}, which suggests that low-frequency observations at large angular separations from the Sun are equally important, for PTAs as well as to study the spatial and temporal structure of IISM variations.

Similar studies with a larger sample of pulsars with existing and upcoming low-frequency facilities could help to improve SW and IISM models. Combining the data from multiple low-frequency observations, which are most sensitive to DM variations, can improve observing cadences and increase bandwidths, allowing for the study of these variations on shorter timescales.  This could potentially provide a detection of the variation in DM, SW activity, or IISM behavior with frequency.

\section{Acknowledgements}
We thank the referee for constructive suggestions. Construction of the LWA
has been supported by the Office of Naval Research under Contract N00014-07-C-0147 and by the
AFOSR. Support for operations and continuing development of the LWA1 is provided by the Air Force
Research Laboratory and the National Science Foundation under grants AST-1835400 and AGS1708855. We thank NOAA's National Centers for Environmental Information (NCEI) for archiving the WSA-ENLIL data used in this study (\url{https://dx.doi.org/10.7289/V5445JGH}).

%%%%%%%%%%%%%%%%%%%%%%%%%%%%%%%%%%%%%%%%%%%%%%%%%%
\section*{Data Availability}
All the pulsar data used in this paper are publicly available on the LWA Pulsar Archive as mentioned in section \ref{sec:observe}. The WSA archive files are publicly available at \url{https://www.ngdc.noaa.gov/enlil/}.



\bibliographystyle{mnras}
\bibliography{pulsardmbib} % if your bibtex file is called example.bib

%\begin{thebibliography}{99}
%\end{thebibliography}

%%%%%%%%%%%%%%%%%%%%%%%%%%%%%%%%%%%%%%%%%%%%%%%%%%

%%%%%%%%%%%%%%%%% APPENDICES %%%%%%%%%%%%%%%%%%%%%

%%%%%%%%%%%%%%%%%%%%%%%%%%%%%%%%%%%%%%%%%%%%%%%%%%

% Don't change these lines
\bsp	% typesetting comment
\label{lastpage}
\end{document}